\documentclass[proceedings]{JHEP} \usepackage{epsfig}
\title{On determining CKM angles $\alpha$ and $\beta$}
\author{Isard Dunietz\\
Fermi National Accelerator Laboratory, P.~O.~Box~500, Batavia, IL
60510-0500, U.S.A.\\
E-mail: \email{dunietz@fnal.gov}}
\conference{Heavy Flavours 8, Southampton, UK, 1999}
\abstract{Because the $B_d \rightarrow J/ \psi K_S$ asymmetry determines
only $\sin{2\beta},$ a discrete ambiguity in the true value of $\beta$
remains.
This note reviews how the ambiguity can be removed. Extractions of the CKM
angle  $\alpha$ are discussed next. Some of the methods require very large
data samples and will not be feasible in the near future.  In the
near future, semi-inclusive CP-violating searches could be undertaken,
which are reviewed last.}
\begin{document}
\section{CKM angle $\beta$}
The primary goal of the various $B$-factories is to test most incisively
the standard CKM (Cabibbo-Kobayashi-Maskawa)~\cite{ckm} description of CP
violation.

For that purpose, the CKM angle $\beta$ extraction via the ``golden'' $B_d
\rightarrow J/ \psi K_S$ asymmetry \cite{bigisanda} can be contrasted to
those via the $B \rightarrow \phi K_S, \eta' K_S, D\overline{D}, 
D_{CP}^0\rho^0,  \mbox{etc.}$ asymmetries. Any significant discrepancy in the
measured $ \beta\ $ values indicates physics beyond the standard
model~\cite{grossmanworah}.

The CKM predictions can be tested more incisively by removing discrete CKM
ambiguities. The CP-violating asymmetry of
$B_d \rightarrow J/ \psi K_S$ allows the determination of $\sin{2
\beta}$~\cite{bigisanda, cpreview}.
A discrete ambiguity in determining $\beta \in [0, 2 \pi)$ remains.
Measuring $ \cos{2\beta}$ removes the ambiguity partially and can be
accomplished,
either by
\begin{description}
\item [(a)] correlating  \(B_s(t) \rightarrow J/ \psi\phi\;\; \mbox{with}\\
B_d(t) \rightarrow J/\psi(\pi^0 K_S)_{K^*} \;
[B_d(t) \rightarrow J/\psi \rho^0] \; \) decays~\cite{ddf}.
\item [(b)] studying the decay-time $(t_K)$ of the  produced neutral
kaon~\cite{azimov,kayserstodolsky} in the process \[B_d(t) \rightarrow
J/\psi
 \stackrel{(-)}{K^0}(t_K),\;\; \stackrel{(-)}{K^0}(t_K) \rightarrow \pi
\ell \nu, \pi \pi . \]
\item [(c)] analyzing Dalitz plots~\cite{charlesetal} of \[B_d(t)
\rightarrow D \overline{D}K_S, D_{CP}^0 \pi^+ \pi^-,  \ldots\]
\item [(d)] using the $B_d-\overline{B}_d$ width difference\footnote{
If  a non-zero width difference $(\Delta\Gamma)_{B_d}$ has been measured,
then
$\cos{2\beta}$ can be obtained from the time-dependence of the untagged
$J/\psi K_S$
sample.
} \[(\Delta\Gamma/ \Gamma)_{B_d} \;\raisebox{-.4ex}{\rlap{$\sim$}} \raisebox{.4ex}{$<$}\;   1\%. \ ~\cite{bbd} \]
\end{description}
Further methods to remove ambiguities can be found in Ref.~\cite{ambiguityremoval}.
\subsection{Physics of Ambiguity Removal}
The underlying reason on how $\cos{2\beta}$ enters is in each case
trivial. For instance,  consider the above method $(d)$.
The interference term $\lambda$ is defined by
\begin{equation}
  \lambda \equiv \frac{q}{p}\frac{<f|\overline{B^0}>}{<f|B^0>} =
-e^{-i2\beta}\ \mbox{ for } f=J/ \psi K_S .
\end{equation}
The  coefficients $q$ and $p$ describe the mass-\-eigen\-states in terms of
$B^0$ and $\overline{B^0}$ states, re\-spec\-tively \cite{cpreview,untagged,annals}.
Note that $\lambda$ is an observable, i.e., a rephase-invariant
quantity~\cite{annals,ddw}. Thus, both $Im \lambda$ and $Re \lambda$ are
measurable, in principle.

The conventional CP-asymmetry measures $Im \lambda$, and is given by (ignoring
$\Delta\Gamma$):
\begin{equation}
\mbox{Asym}(B_d(t) \rightarrow J/\psi K_S) = -Im \lambda\;\;  \sin{\Delta mt}.
\end{equation}
However, $Re \lambda$ enters in the untagged
time-depen\-dence~\cite{untagged},\footnote{
Eq.~(1.3) is correct for $|q/p|=1$, which holds to an excellent
approximation within the CKM model. However, when the statistics
gets sufficiently large to detect effects due to a non-vanishing width
difference, then also the effects due to $|q/p|\neq
1$~\cite{altomari} may have to
be incorporated.
Determining the sign of $\cos{2\beta}$ from Eq.~(1.3) requires independent
knowledge of whether $\Delta\Gamma \equiv \Gamma_H - \Gamma_L$ is positive
or negative.
This independent knowledge may be very difficult to achieve. Thus, the
argument
can be reversed and Eq.~(1.3) may be used to determine the observable 
sign($\Delta\Gamma$),  because $\mbox{sign} (\cos{2\beta})$ will be known by
other means.
}

\begin{eqnarray}
\Gamma[J/ \psi K_S(t)]  &\equiv&  \Gamma(B_d(t) \rightarrow J/ \psi K_S) + \nonumber
\nonumber\\
                      & & \Gamma(\overline{B_d}(t) \rightarrow J/ \psi
K_S)  \nonumber \\
                    &\sim & e^{- \Gamma _Lt}+ e^{- \Gamma _Ht} +  \nonumber
\\
                    &  &Re \lambda(e^{-\Gamma _Lt}- e^{- \Gamma _Ht}).
\end{eqnarray}
Because the expected  $\Delta\Gamma/\Gamma$ is tiny,  an excess of $10^5$\
untagged $J/ \psi K_S$ events is required.  Then studies of effects
dependent on the $B_d-\overline{B_d}$ width difference become feasible.

While the above discussion may become relevant only in the far future of
$B_d$ physics,  it is of more immediate importance for $B_s$ physics.

The $B_s-\overline{B_s}$ width difference is predicted to be sizable
(around 10\%)~\cite{bbd},  and once observed will permit the
\underline{unambiguous}~\cite{untagged}
extraction of CKM phases in $B_s(t) \rightarrow f$ processes. For example,
a time dependent study of  $B_s(t) \rightarrow D_s^\pm K^\mp$~\cite{adk} 
will determine the CKM angle $\gamma$ unambiguously. Experiments
where $B_s$ mesons are copiously produced, may be able to make extensive
use of this opportunity.

\section{CKM angle $\alpha$}
The CKM matrix can be completely specified by four independent quantities.
The three angles of the CKM unitarity triangle satisfy
$$ \alpha =\pi -\beta -\gamma , $$
and thus are not independent.

Since we were asked to discuss the extraction of the angle $\alpha$, we
should have reviewed the determination of the CKM angle $\gamma$.
The angle $\gamma$ can be determined from
\begin{description}
\item [(a)] a $B \rightarrow K\pi$
analysis~\cite{neubertRosner},
\item [(b)] $B_s(t) \rightarrow D_s^\pm K^\mp$
studies~\cite{adk},
\item [(c)] a $B^- \rightarrow D^0K^-,  \overline{D^0}K^-$
analysis~\cite{ads},
\item [(d)] Dalitz plot analyses~\cite{enomototanabashi},
\item [(e)] $B_d(t) \rightarrow \pi ^+\pi ^-$ and $ B_s(t) \rightarrow
K^+K^-$ correlations~\cite{snowmass,fleischerGamma},
\item [(f)] $ B_d(t) \rightarrow J/\psi K_S$ and $B_s(t) \rightarrow
J/\psi K_S$ correlations~\cite{snowmass,fleischerGamma}.
\end{description}
However,  Neubert addressed the extraction of the CKM angle
$\gamma$~\cite{neubert}, and this note thus reviews the ``traditional" CKM
$\alpha$ determinations.

The angle $\alpha$ can be determined from
\begin{description}
\item [(1)] the $B_d(t) \rightarrow \pi ^+\pi ^-$ asymmetry if penguin
amplitudes were negligible,
\item [(2)] $B_d(t) \rightarrow \rho \pi$ Dalitz plot
analyses~\cite{snyderQuinn},
\item [(3)] $B_d(t) \rightarrow D^{(*)^\pm} \pi ^\mp$ studies~\cite{dpi}.
\end{description}
Penguin amplitudes in the $B \rightarrow \pi ^+\pi^-$ process are likely to be
 sizable,
 as can be inferred from the recent CLEO measurement~\cite{cleo}
\[ \frac {B(B \rightarrow K\pi)}{B(B \rightarrow \pi\pi)} \approx 4, \]
and the naive approach (1) will probably not work.
 The CKM angle $\alpha$ can be
extracted by selecting the ``penguin-free" $B \rightarrow (\pi\pi)_{I=2}$
process~\cite{gronaulondon}.\footnote{
Electro-weak penguin amplitudes may have to be accounted
for also~\cite{desh}.
}
The selection requires studies of $B \rightarrow \pi
^0\pi ^0$,
which is not feasible with first generation $B$-factory experiments.

However, recent theoretical advances indicate that it may be possible to
determine the CKM angle $ \alpha$ from the $B_d(t) \rightarrow \pi ^+\pi
^-$
asymmetry alone~\cite{benekebns}.

Approach (2) requires large statistics~\cite{lediberder}. But once
obtained,  the CKM angle $\alpha$ can be extracted even if penguins are
present.
Electro-weak penguin contributions may introduce sizable uncertainties,
and must be studied further.\footnote{
They were found to be small in particular models~\cite{charles}.
}

The $D^{(*)^\pm} \pi ^\mp$ processes permit the clean determination of
$\beta - \alpha$ or of $2\beta +\gamma$ because no penguins are involved
~\cite{dpi}. Since $\beta$ will be known $\alpha\ $  (or $\gamma)$ can
thus be determined.
\section{Near Future}
For the near future,  experiments will not be sensitive to CP violationg
effects with tiny branching ratios,  because of
limited integrated luminosity. One may still be able to study
(semi-) inclusive CP asymmetries.

For instance, mixing-induced CP violation can be searched for in double
charm, single charm and charmless $B^0$ samples~\cite{inclusiveCP,inclusiveCPIsi,bbd}.

\TABULAR{|l|c|}{\hline
{Final state of $B^0(\equiv B_d^0\ \mbox{or}\  B_s^0)$} & Required number of flavor tagged\\
& $B_s^0\ \&\  \overline {B_s^0}\ \ (B_d^0\ \&\  \overline{B_d^0}$) mesons\\\hline
{double charm}      & {$2 \times10^5\ (2\times 10^6)$} \\\hline
{single charm}       & {$6 \times10^5\ (8 \times 10^6)$}\\\hline
{charmless}        & {$ 10^6\ (2\times 10^7)$}\\\hline}{Number 
of flavor-tagged $B^0$ plus $ \overline{B^0}$
mesons necessary to observe a $3\sigma$ CP-violating
effect (column 2) in the modes specified by column 1.}

Table I lists the required number of tagged $B^0$ and $\overline{B^0}$
mesons to observe $3\sigma$ effects~\cite{inclusiveCPIsi}.
Such effects, once observed, can be related to 
CKM parameters~\cite{inclusiveCP,inclusiveCPIsi,bbd}.

Other promising mixing-induced CP asymmetries are
\begin{description}
\item [(1)] $B^0(t) \rightarrow J/\psi X \; $     versus
$\; \overline{B^0}(t) \rightarrow J/\psi X, \\$
\item [(2)] $B^0(t)  \rightarrow \mbox{(primary $K_S$)} X$ \ versus\ \\
$\overline{B^0}(t) \rightarrow \mbox{(primary
$K_S$)}X.$ 
\end{description}

All the above effects in this Section require flavor-tagging,  which is expensive. The
flavor-tagging requirement reduces the statistical reach by an order of
magnitude.

Thus, direct CP violation should be searched for also. It requires neither
flavor-tagging nor mixing nor time-dependences. [At hadron colliders,
the long b-lifetimes are a blessing and provide the primary distinction
between b-hadrons and backgrounds. For hadron colliders,
time-dependences are no hindrance.]

Browder et al.~\cite{browder} suggested to search for semi-inclusive CP
asymmetries in
\[B \rightarrow K^+X,  K^*X \;\;   \mbox{versus}\;\; \overline{B}
\rightarrow K^-X, \overline{K^*}X, \]
where the $K^{(*)}$ has a very high momentum. The $BR \sim 10^{-4}$ and
the CP asymmetries are expected to be $\;\raisebox{-.4ex}{\rlap{$\sim$}} \raisebox{.4ex}{$<$}\; 10\%$. Additional semi-inclusive CP-violating effects were
discussed in Ref.~\cite{moreInclCP}.

The semi-inclusive $b \rightarrow J/\psi +d$ processes also may
exhibit direct CP asymmetries at the $ \;\raisebox{-.4ex}{\rlap{$\sim$}} \raisebox{.4ex}{$<$}\; \mbox{few}\, \%$
level~\cite{psicp,snowmass}.
Their $BR \approx 5 \times 10^{-4}$ and the effect can be searched for in charged $B^\pm$ decays,
\[N(J/\psi X_d^+) \neq N(J/\psi X_d^-).\]
\section{Conclusions}
The CKM quantity $\sin{2\beta}$ will soon be measured accurately from the
$B \rightarrow J/\psi K_S$ asymmetry.

Measurements of the sign of $\cos{2\beta}$ will test the CKM model more
incisively (see Section 1). Section~1 emphasizes that time-dependent
studies of $B_s$ decays can determine CKM parameters without any
discrete ambiguity!
The relevant $B_s$ modes thus probe the CKM model in detail.

Section~2 discusses several ways of determining the CKM angle $\alpha$.
Because CP effects with tiny branching ratios are unreachable in the near
future, Section~3 suggests several (semi-)inclusive CP asymmetries,
some of which could even yield valuable CKM information.

\acknowledgments
I thank the organizers and their 
excellent staff for making the Heavy Flavours 8 conference at Southampton 
so 
successful and pleasant.  I am grateful to G. Buchalla, R. Kutschke and 
U. Nierste for proofreading this manuscript.

\end{document}